# Design and Analysis of a Controller Using Quantitative Feedback Theory for a Vehicle Air Suspension System

A. Shafiekhani, S.M. Mirsadeghi, K. Torabi
Mechanical Engineering Department
University of Kashan

*Abstract*—This paper presents the design of a robust controller for a vehicle air suspension system using Quantitative Feedback Theory (QFT). This study is primarily focused on control of linearized active air suspension system. For the purpose of simplicity, the dynamics of the air suspension system is modeled using a simple 2-DOF quarter car model. Uncertain dynamic system with different working condition has been considered for the vehicle air suspension system.

## I. INTRODUCTION

All vehicle suspension are designed with the same target i.e., to filter out vibration coming from the tire in contact with road and contributing to the handling of the vehicle. Air suspensions have some features, which are not easy to obtain with the mechanical suspensions [1]. Variability of the ride height, reduced weight, adjustable carrying capacity, and reduced structurally transmitted noise are the main advantages of the air suspensions over the conventional mechanical ones [2]. The design and optimization of air suspension systems have been reported during recent years. Quaglia et al, introduced an improved one DOF quarter model for modeling the air suspension systems [3]. The non-linear equations of motion governing the dynamic behavior of the 2-DOF pneumatic quarter car suspension model were presented by Vogel [4]. Porumamilla obtained a linear model for this pneumatic suspension [5].

Considering a suitable control system to accomplish performance specifications in the presence of uncertainties (plant changes and external disturbances) is a key point in any feedback design [6, 7]. There are many practical systems that have high uncertainty levels in their open-loop transfer functions which make it very difficult to create appropriate stability margins and good performance in command following problems for a closed-loop system [8–10]. The modeling of control system has been used from the models proposed in Yazdanpanah et al. works [11]. Also fuzzy control systems can be used to enhance the performance of the control structure [12]. Therefore, a single fixed controller in such systems is found among the robust control family. Quantitative Feedback Theory (QFT) is a robust feedback control system design technique which allows the direct design to closed-loop robust performance and stability specifications [13–15]. Based on QFT, one of the main objectives is to design a simple, low-order controller with minimum bandwidth. Many of the techniques applied to the robust control family such as $H\infty$ are based on the magnitude of a transfer function in the frequency domain. QFT not only uses this transfer function approach but also takes into account phase information in the design process. The unique feature of QFT is that the performance specifications are expressed as bounds on the frequency-domain response. Meeting these bounds implies a corresponding approximate closed-loop realization of the time-domain response bounds for a given class of inputs and for all uncertainty levels in a given compact set. These suspension systems can be implemented in industrial arm robots [16].

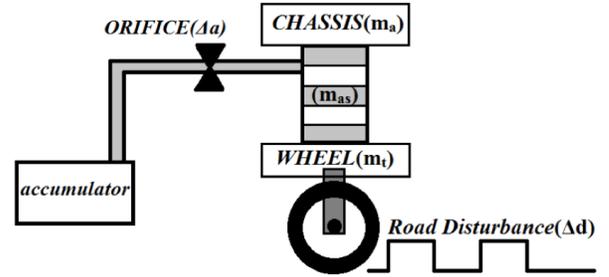

Figure 1. Schematic model of the 2-DOF air suspension system

## II. FORMULATION

The pneumatic processes that govern the performance of the isolator are inherently nonlinear and irreversible. The fidelity of the pneumatic system model depends on the extent of the computational complexity in the algorithm which captures the nonlinear mechanical and thermodynamic behavior.

Linearized state space model of the 2-DOF pneumatic quarter car suspension system are taken from [5] and are as given by equation below:

$[x_1, x_2, x_3, x_4, x_5] = [x_a, \dot{x}_a, x_t, \dot{x}_t, m_{as}]$

where, $x_a$ is absolute displacement of the chassis, $\dot{x}_a$ is absolute velocity of the chassis, $x_t$ is absolute displacement of the wheel, $\dot{x}_t$ is absolute velocity of the wheel and $m_{as}$ is mass of air in the air spring.

$\triangle \dot{x} = [A] \triangle x + [B_u] \triangle a + [B_d] \triangle d$
$y = [C] \triangle x + [D_u] \triangle a + [D_d] \triangle d$

Table I
NOMINAL VALUES OF THE SYSTEM PARAMETERS

| Q1 = -175 | Q2 = 6905 |
|---|---|
| P1 = 1486 | P2 = 58720 |
| S1 = 60.162 | S2 = 2380 |
| | S3 = 51 |

Table II
NOMINAL AND PERTURBED PARAMETRIC VALUES FOR THE PNEUMATIC MODEL CHASSIS TIRE

| Ma =90  10 kg | Mt=16  5 kg |
|---|---|
| Ka =implicit in Eq. of motion | K= 1e5  0.1e2 N/m |
| Ca=50  10 Ns/m | Ct =600  100 Ns/m |

$$A = \begin{bmatrix} 0 & 1 & 0 & 0 & 0 \\ Q_1 & -\left(\frac{C_a}{m_a}\right) & -Q_1 & \left(\frac{C_a}{m_a}\right) & Q_2 \\ 0 & 0 & 0 & 1 & 0 \\ P_1 & \frac{C_a}{m_t} & -\left(P_1+\frac{K_t}{m_t}\right) & -\left(\frac{C_t+C_a}{m_t}\right) & P_2 \\ S_1 & 0 & -S_1 & 0 & -S_2 \end{bmatrix}$$

$B_u = \begin{bmatrix} 0 & 0 & 0 & 0 & 0 & S_3 \end{bmatrix}^T$

$B_d = \begin{bmatrix} 0 & 0 & 0 & \frac{K_t}{m_t} & 0 \end{bmatrix}^T$

$C = \begin{bmatrix} 1 & 0 & 0 & 0 & 0 \end{bmatrix}$

$D_u = 0, D_d = 0$

The control input to the system is the orifice area $\Delta a$ and the disturbance input affecting the system is in the form of the road displacement $\Delta d$. Absolute displacement of the chassis is output of model which is measured and used as the feedback of the control system.

The parametric values required to generate the state space matrices are given in tables I and II.

Using this sate space model, transfer function of the air suspension system with uncertain parameters can be obtained as shown below:

$G_d(S, \alpha_i) = \frac{(a_1 S^4 + a_2 S^3 + a_3 S^2 + a_4 S + a_5)}{(S^5 + b_1 S^4 + b_2 S^3 + b_3 S^2 + b_4 S + b_5)}$

$G_u(S, \alpha_i) = \frac{c_1 S^4 + c_2 S^3 + c_3 S^2 + c_4 S + c_5}{S^5 + b_1 S^4 + b_2 S^3 + b_3 S^2 + b_4 S + b_5}$

where,

$a_1 = [-1.82, 3.64]10^{-12}$
$a_2 = [-2.47, 3.49]10^{-10}$
$a_3 = [2500, 4688]$
$a_4 = [7.04, 12.25]10^6$
$a_5 = 6.7610^6$
$b_1 = [2414, 2428]$
$b_2 = [8.92, 12.28]10^4$
$b_3 = [21.98, 22.04]10^6$
$b_4 = [7.08, 12.3]10^6$
$b_5 = 6.7610^6$
$c_1 = [-3.64, 3.18]10^{-12}$
$c_2 = [-1.89, 4.08]10^{-10}$
$c_3 = 3.5210^5$
$c_4 = [1.31, 1.90]10^7$
$c_5 = 3.2510^9$

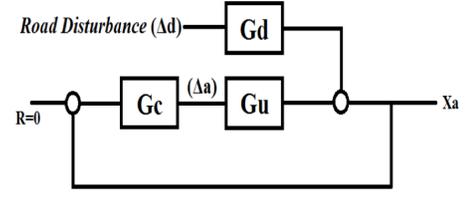

Figure 2. Block diagram of the closed loop system

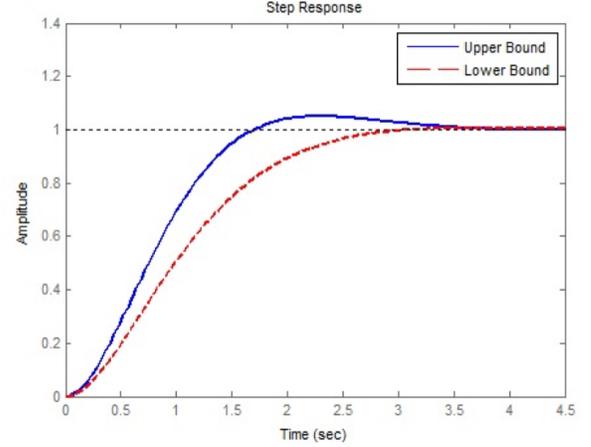

Figure 3. The desired time response bounds of the system

## III. ROBUST CONTROL TRACKING MODELS

The QFT approach for tracking the reference has been implemented in this section. Based on the block diagram of the system, the tracking condition should be met as below:

$|\frac{GPH}{(1+GPH)}| \leq W_s t,\ for\ all\ P \in P, \omega \in [0, \infty)$
$W_s t = 1.2$
$\begin{cases} G = G_c \\ P = G_u \\ H = 1 \end{cases}$

where $W_s t$ is computed by considering 5% overshoot for the upper bound. Also the desired settling time for upper and lower bounds and the appropriate rise time for the lower bound for the suspension system is assumed to be 3 and 1.7 seconds respectively. Using these characteristics of the system, the time response and bode diagram of the system is plotted as shown in Figures 3 and 4.

where, $\delta = 20 log(W_s t)$

Following the procedure the required bounds for tracking model has been plotted and the frequency response of the uncertain plant is obtained for frequencies included in the frequency array which is chosen based on the performance bandwidth.

## IV. DISTURBANCE REJECTION MODELS

Based on the desired characteristics of the disturbance rejection model, has been chosen such that it satisfies the performance expected by the designer.

$|\frac{(A+BG)}{(C+DG)}| \leq W_s d$

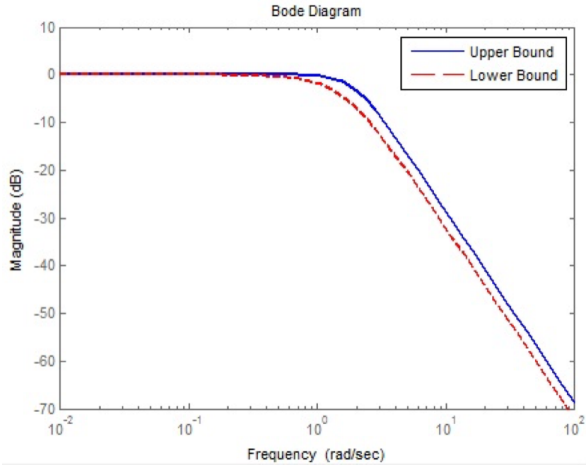

Figure 4. Bode diagrams of the desired system characteristics

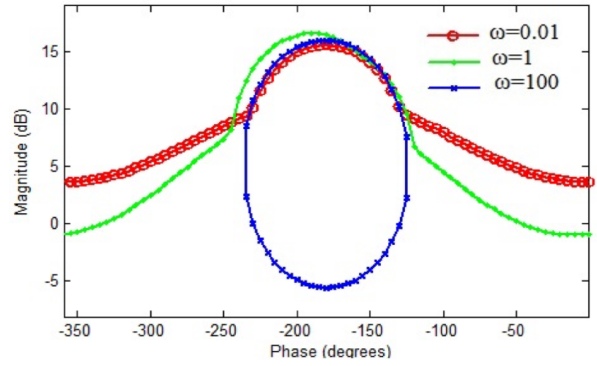

Figure 6. Intersection (worst-case) of all bounds

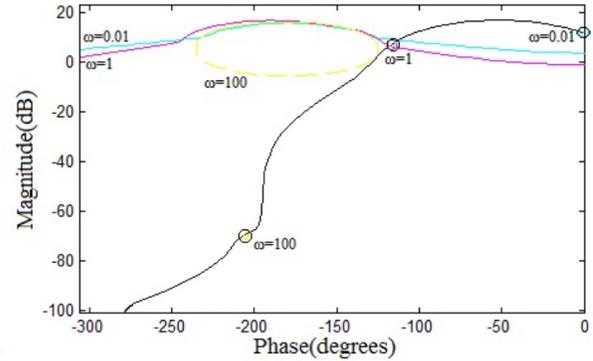

Figure 7. Final design with QFT procedure

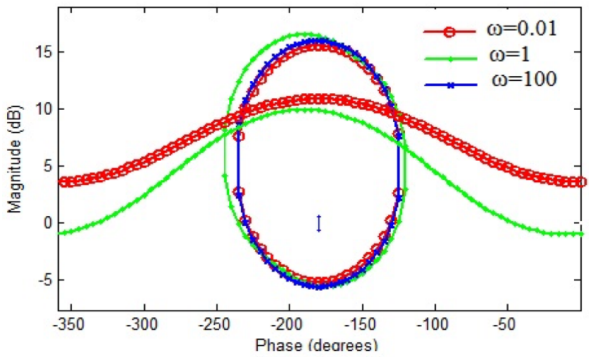

Figure 5. Superposition of all bounds

$$W_s d = 0.4$$
$$\begin{cases} A = G_d \\ B = [0] \\ C = [1] \\ D = Gu \\ G = Gc \end{cases}$$

## V. CONTROLLER DESIGN

In order to design the controller, The QFT bounds must be computed for tracking and disturbance rejection separately by employing the QFT controller design toolbox. Thus, the frequency response sets for tracking and disturbance rejection models have been plotted using sisobnd6 and genbnds10 respectively. Figure 5 shows both models in different frequencies. Following QFT controller design procedure, the intersection of each model in different frequencies has been depicted in Figure 6. The frequency array has been chosen based on the performance bandwidth and shape of the templates.

Considering the mentioned bounds, the next step to follow in a QFT design procedure is called loop shaping which means designing of a nominal loop function that meet the bounds. The nominal loop is constructed by product of the nominal plant and the controller which has to satisfy the worst case of the bounds. The proper controller has been designed by adding required elements as below:

- Two real Poles on -9.45 and -4.3
- Two real Zeros on -0.84 and -20.2
- A complex pole with Re=-309.6 and Im=309.7

The transfer function of the controller has been computed as:
$G_c = \frac{(3673s^2 + 7.729 \times 10^4 s + 6.233 \times 10^4)}{(s^4 + 632.9s^3 + 2.003 \times 10^5 s^2 + 2.662 \times 10^6 s + 7.791 \times 10^6)}$

The final results of the loop shaping step has been shown in Figure 7.

## VI. PERFORMANCE ANALYSIS

After completing the QFT design, the response of the closed loop system must be analyzed. Therefore, the computed displacement and acceleration responses of the nominal closed loop and open loop systems subjected to various road disturbances have been plotted in figures 8 and 9.

At first, the road disturbance which may be in the form of two bumps is modeled as two input pulses with height of 5 cm.

Secondly, the road disturbance has been supposed to be as an impulse. The resulted responses for open loop and closed loop have been compared in Figure 10 and 11.

Finally, Figure 12 and 13 show the chassis displacement and acceleration in which the road disturbance has been modeled using a white noise.

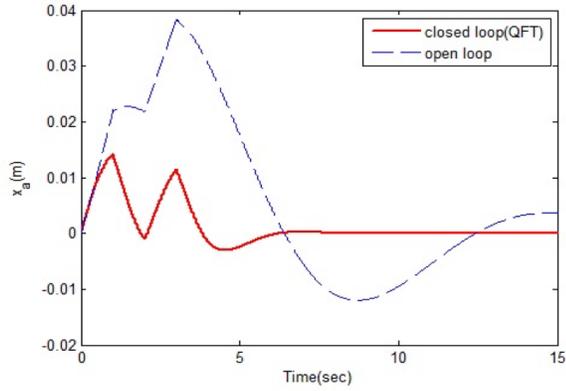

Figure 8. Chassis displacement for two bumps as an input

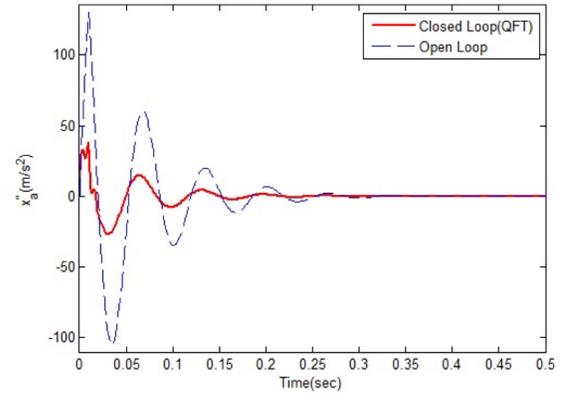

Figure 11. Chassis acceleration for impulse as an input

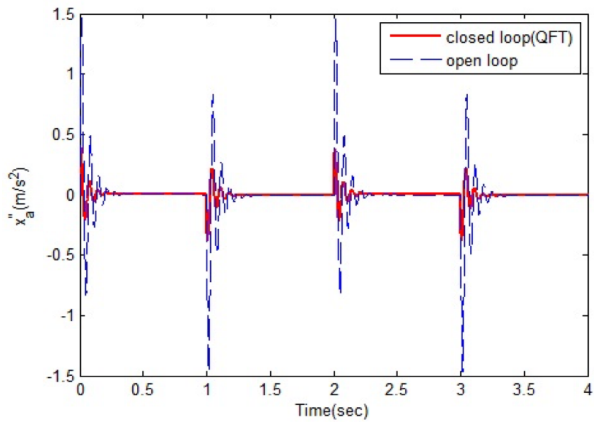

Figure 9. Chassis acceleration for two bumps as an input

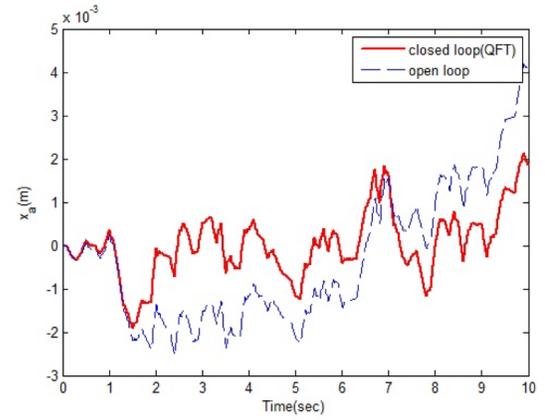

Figure 12. Chassis displacement for white noise as an input

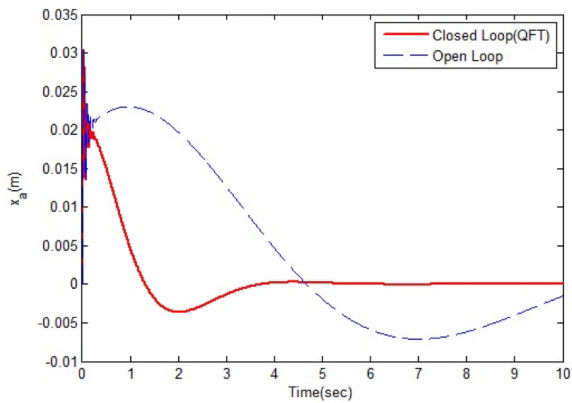

Figure 10. Chassis displacement for impulse as an input

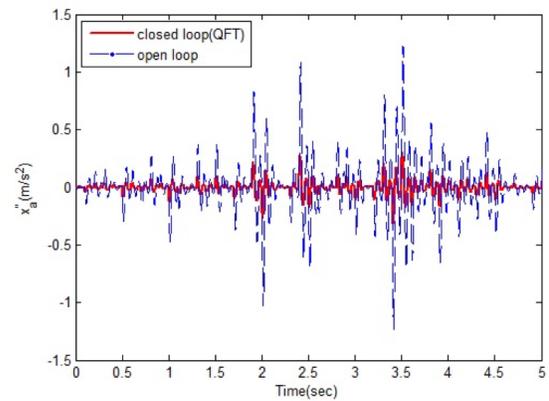

Figure 13. Chassis acceleration for white noise as an input

## VII. CONCLUSION

In this paper, a robust controller using QFT for a vehicle air suspension system was designed and evaluated. Since there are many uncertainties in the mathematical model of an air suspension system, the QFT was chosen for designing the controller to improve the suspension performance. The procedure of desinging the QFT controller was explained and the proposed controller was implemented on a 2-DOF model of air suspension system. Results clearly showed that the controller is improving the performance in exsistance of different disturbance inputs.